\documentstyle[12pt,floats,aps]{revtex}
\include{psfig}
\newcommand{\mywidth}{3.4in}

\begin{document}
\draft

\title{Scaling of the specific heat of superfluids confined in pores}
\author{Norbert Schultka$^1$ and Efstratios Manousakis$^2$}
\address{$^1$ Institut f\"ur Theoretische Physik, Technische
Hochschule Aachen, D-52056 Aachen, Germany\\ $^2$Department of Physics, Florida State University, Tallahassee,
Florida 32306, USA}
\date{\today}
\maketitle
\begin{abstract}
We investigate the scaling properties of the specific heat of the XY model 
on lattices $H \times H \times L$ with $L \gg H$ 
(i.e. in a bar-like geometry) with respect to the thickness $H$ of the bar, 
using the Cluster Monte Carlo method. We study the effect of the geometry and
boundary conditions on the shape of the universal scaling function of
the specific heat by comparing the scaling functions obtained for cubic,
film, and bar-like geometry. In the presence of physical boundary
conditions applied along the sides of the bars we find good agreement between 
our Monte Carlo results and the most recent experimental data 
for superfluid helium confined in pores.
\end{abstract}
Bulk liquid $^4$He exhibits a second order phase transition at 
the $\lambda$-critical 
temperature $T_{\lambda} \approx 2.18$K where it turns superfluid. 
Approaching  $T_{\lambda}$ from below the specific heat $c$ and the 
superfluid density 
have singularities which are characterized by the critical exponents 
$\alpha$ and $\nu$, respectively; these critical indices determine the 
universality class of this second order phase transition. If liquid 
$^4$He is placed in a confining geometry, e.g. film or pore-like
geometry, the finite-size scaling theory\cite{FISHER} 
can be used to describe the behavior of the physical quantities
at temperatures close to $T_{\lambda}$. The finite-size scaling theory
is based on the assumption that the system feels its finite size 
when the correlation length $\xi$ becomes of the order of the  
confining length. For a physical quantity $O$ this statement can be 
expressed as follows\cite{BREZIN}:
\begin{eqnarray}
\frac{O(t,H)}{O(t,H=\infty)} &=& f(x),\\
 \label{opfss}
\frac{H}{\xi(t,H=\infty)} &=& x,
\end{eqnarray}
where $H$ denotes the relevant confining length and the reduced temperature
$t=T/T_{\lambda}-1$ while $\xi(t,H=\infty)$ is the correlation
length of the bulk system. The point is  that the
dimensionless function $f$ depends only on the
dimensionless ratio $H/\xi$ and it does not depend on microscopic
details of the system. It does, however, depend on the observable $O$,
the type of confining geometry and on the conditions imposed (or not, in
the case of free boundaries) at the boundaries of the system.

Liquid $^4$He can be an ideal testing ground to check 
the validity of the finite-size scaling theory experimentally\cite{LIPA}
because the specific heat 
$c$ and the superfluid density can be measured to a very high accuracy.
In addition, the shape of the confining geometry, such as films or
pores, can be designed with such a precision that the relevant 
confining length is well defined. 

In order to compare the results of finite-size scaling theory with the 
experimental results on liquid $^4$He, there have been theoretical 
efforts to compute the universal
finite-size scaling function of the specific heat of confined liquid 
$^4$He\cite{SWDF,DOHM,RGDIB,EPSI,SCHU1,SCHU2,SCHU3}. 
The scaling function for film geometry has been computed by 
loop expansion based renormalization group treatment of
the standard $\phi^4$ Landau-Ginzburg functional\cite{SWDF,DOHM,RGDIB,EPSI} 
and by the Monte-Carlo method within the XY 
model\cite{SCHU1,SCHU2,SCHU3}. The XY model is another
form of the standard Landau-Ginzburg functional and both models belong
to the same universality class as liquid $^4$He, i.e. all three systems 
have the same critical indices. The scaling functions for a film 
geometry have been computed from the XY model using the Monte Carlo
method\cite{SCHU1,SCHU2,SCHU3} and they 
are in rather good agreement with the scaling functions obtained from
the field theoretical treatment of the $\phi^4$ Landau-Ginzburg 
theory\cite{RGDIB}. 
The scaling functions obtained when 
periodic boundary conditions\cite{SCHU2}  were applied on the top
and on the bottom boundary of the film are very different from the
scaling functions obtained when staggered boundary 
conditions\cite{SCHU1,SCHU3} (vanishing order parameter on the boundary
layer) were applied on the top and bottom film
boundary. The scaling functions obtained
by the Monte Carlo simulation of the XY model\cite{SCHU1,SCHU3}
agree reasonably well with the experimentally determined\cite{LIPA}
scaling functions when staggered boundary conditions 
were applied in the simulation.
Such boundary conditions on the order parameter  are believed to 
approximate better the real conditions on the top and bottom boundary of 
superfluid film\cite{RGDIB,SCHU1}. 

Besides the boundary conditions, the other important factor which,
in principle, can determine the scaling function is the geometry. 
Close enough to the critical point
where the bulk correlation length $\xi$ 
becomes of the size of the confining length $H$ or even
larger, the system as a whole ``knows'' about the geometrical shape of the 
confining volume. Thus, one expects to find a region of the
dimensionless variable 
$H/\xi$ where the form of the scaling function is sensitive to the
geometry of the confining volume.  There are experiments which have
been performed for the pore geometry\cite{EARLYC,CHEN,COLEMAN} and the 
results for the specific scaling function are significantly different 
from those obtained from the film geometry.
In this paper, we report results of our simulations of the XY model
in a bar-like geometry, namely on lattices $H \times H \times L$ with
$L  >> H$, which mimic the pore-like geometry of the experiments. In
addition, to represent the real situation more closely we have used 
staggered boundary conditions along the confining dimensions ($H$
directions).  Periodic boundary conditions are used along the long ($L$)
direction of the bar because they approximate the limit $L \to \infty$ better.
We find that our results for the scaling function are different from those
obtained for the film geometry and  are also in reasonably good
agreement with the most recent experimental results\cite{COLEMAN}
without any adjustable parameter.

In order to describe the fluctuations of the 
order parameter in superfluid $^{4}He$ near the $\lambda$ 
point we employ the XY model which belongs to the same universality
class as the Landau-Ginzburg free energy which is expressed in terms of
the superfluid order parameter $\psi(\vec r)$. Within the 
XY model the free-energy $\cal H$ can be written as
\begin{equation}
{\cal H} = -J \sum_{\langle i,j \rangle} \vec{s}_{i} \cdot 
     \vec{s}_{j}, \label{ham}
\end{equation}
where the summation is over all nearest neighbors,
$\vec s = (\cos\theta, \sin\theta),$ and $J>0$ sets the energy
scale.   In this model $\vec s_i$  are not real spins but they are 
pseudospin variables\cite{KLEINERT} in which the
angle $\theta$ corresponds to the phase of the superfluid order
parameter $\psi(\vec r)$. The order parameter can be
understood\cite{KLEINERT} as the average value of operator which 
creates an atom in the superfluid
helium and it is defined in a volume
whose linear extensions are much larger than the interparticle spacing and
much smaller than the correlation length, a condition that is
realized only very near the transition temperature.

The pore geometry is represented by  $H \times H \times L$ lattices with
$L \gg H$, namely, bar-like geometry. On the ends
of the bar we always use periodic
boundary conditions whereas we have carried out separate calculations
for a) periodic and b) staggered boundary conditions 
along the sides ($H$-directions) of the bar. 
In the case of staggered boundary conditions the four sides 
of the bar are coupled
to a staggered pseudospin configuration,
i.e. the pseudospins at the boundaries alternate when we move along 
the lattice bonds (cf. also Ref.\cite{SCHU1}) so that the
total sum of the spins at the sides of the bars is exactly zero.
This boundary condition plays the role of the confining walls  
and corresponds to a boundary condition
where the superfluid order parameter is exactly zero at the boundaries.
In the simulation this can be easily visualized by corresponding every
spin of the boundary layer to the same pseudospin but which interacts with 
all other spins of the layer next to the boundary via a coupling which 
alternates in sign. This spin is a dynamical variable which means 
that it can change direction during the cluster updating procedure and thus, 
the $O(2)$ invariance of the model is not broken at the boundaries.

We computed the specific heat on $H \times H \times L$ lattices, where 
$H=16,20,26,32$ and $L=5H$. The specific heat $c$ is obtained by
\begin {equation}
 c = \frac{\beta^{2}}{N} 
\left( \langle {\cal H}^{2} \rangle - \langle {\cal H} \rangle ^{2} \right),
\label{heat}
\end{equation}
where $\beta = 1/(k_B T)$
and $N$ is the number of pseudospins contributing to the specific heat.
The multi-dimensional integrals in the  expression
for the averages in Eq.(\ref{heat}) were computed by means of the Monte Carlo 
method using Wolff's 1-cluster algorithm \cite{WOLFF1}.
To thermalize the system, we typically carried out of the order of 
$20,000$ steps involving single cluster flips during which no observable was
calculated. After the end of that process we typically carried out
of the order of $750,000$ single cluster flips  
where observables were calculated. The calculations were performed on a 
heterogeneous environment of computers including IBM RS/6000 and DEC alpha AXP
workstations and a Cray Y-MP.

\begin{figure}[htp] 
 \centerline{\psfig {figure=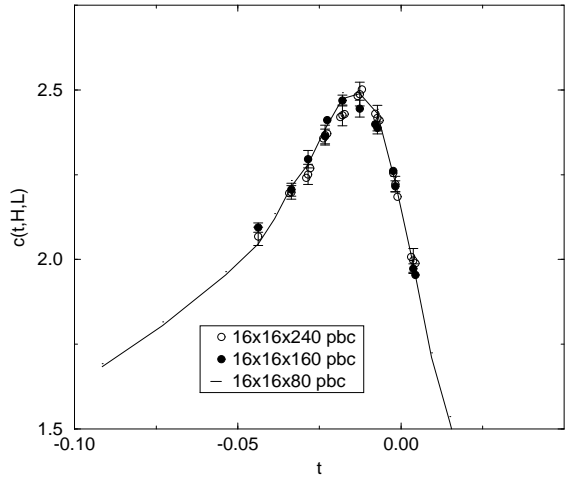,width=\mywidth}}
 \caption{\label{fig1} The specific heat as a function of 
  $t=T/T_{\lambda}-1$ for $16^{2}\times L$ 
lattices with periodic
  boundary conditions (pbc) in all directions. In the 
  three-dimensional XY model $T_{\lambda}/J =2.2017 $\protect\cite{JANTL}.}
\end{figure}
Here we check the finite-size scaling hypothesis for the specific heat
with respect to the bar diameter $H$ in the $L \rightarrow \infty$
limit.  We do not need to take the actual $L \rightarrow \infty$ limit, because
it turns out that lattices with the ratio
$L/H=5$ are a satisfactory approximation to that limit.
We demonstrate this in Fig.~\ref{fig1} where the specific heat,
calculated on lattices with ratios $L/H=5,10,15$ and $H=16$,
is plotted near the peak. Within error bars the results for the three 
size lattices are the same.

We are now able to compute the finite-size scaling function $f_{1}(tH^{1/\nu})$
defined by the expression\cite{SWDF,DOHM}:
\begin{equation}
   c(t,H) = c(t_{0},\infty) + H^{\alpha/\nu } f_{1}(tH^{1/\nu}).
  \label{cimph}
\end{equation}
The function $f_{1}(x)$ is universal and $\nu=0.6705$ as has been
extracted from recent experiments\cite{GOLAHL}. The
hyperscaling relation $\alpha=2-3\nu$ yields $\alpha/\nu=-0.0172$.
At the reduced temperature $t_{0}$ the correlation length
$\xi(t)= \xi_{0}^{\pm} |t|^{-\nu}$ becomes equal to the bar thickness
$H$, i.e. $t_{0}=(\xi_{0}^{+}/H)^{1/\nu}$ with $\xi_{0}^{+}=0.498$ 
\cite{GOTTLOB}. We have
\begin{equation}
c(t_{0},\infty)=c(0,\infty)+\tilde{c}^{+}_{1} t_0^{-\alpha},
\label{c0l}
\end{equation}
where we use the bulk values $c(0,\infty)=30$, 
$\tilde{c}_{1}^{+}=-30$ obtained by studying the finite-size scaling 
of the specific heat of cubes\cite{SCHU2}.

\begin{figure}[htp] 
 \centerline{\psfig {figure=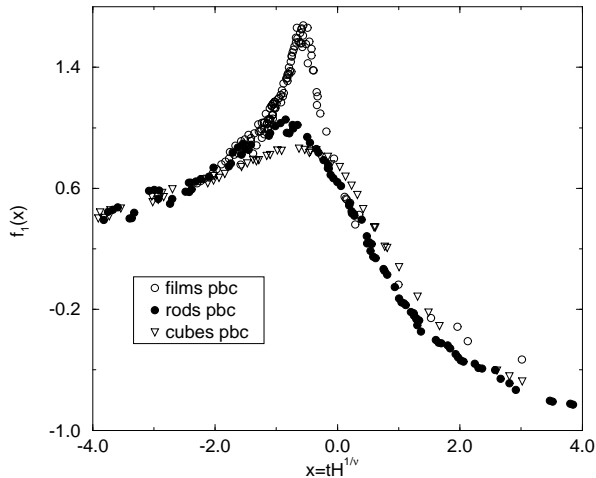,width=\mywidth}}
 \caption{\label{fig2} Comparison of the finite-size scaling functions 
$f_{1}(tH^{1/\nu})$ in Eq.(\protect\ref{cimph})
for films, bars and cubes with periodic boundary conditions (pbc) in all 
directions.}
\end{figure}
In Fig.~\ref{fig2} we compare the scaling functions $f_{1}(x)$ for cubes,
films (taken from Ref.\cite{SCHU2}) and bars in the presence of periodic
boundary conditions. The more confining dimensions occur in the system, i.e.
one, two, and three confining dimensions for the cube, the bar and the 
film, respectively, the more the
function $f_{1}(x)$ is suppressed, only in the limit $|x| \rightarrow \infty$
do the three functions agree.

\begin{figure}[htp] 
 \centerline{\psfig {figure=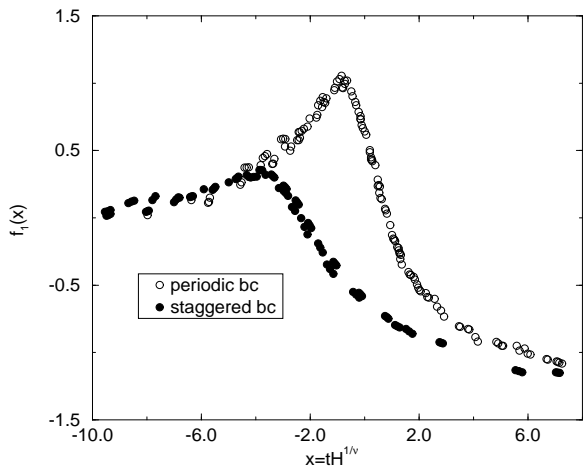,width=\mywidth}}
 \caption{\label{fig3} Comparison of the scaling functions
$f_{1}(x)$ for bars with periodic and staggered boundary conditions on the
sides.} 
\end{figure}
Here we compute the scaling function $f_{1}(x)$ again but using 
staggered boundary conditions in the $H$-directions of the lattice. 
These boundary conditions seem to mimic the physical boundary conditions met 
in the experiments rather well \cite{DOHM,RGDIB,SCHU1}.
Our results for the scaling function $f_1$ are shown in Fig.~\ref{fig3}
where $f_1$ is compared to that obtained with periodic boundary
conditions. This figure demonstrates
that in the presence of staggered boundary conditions the scaling function 
$f_{1}(x)$ is dramatically suppressed compared to the function $f_{1}(x)$
obtained for bar-like geometry with periodic boundary conditions 
in all directions.
For large values of $|x|$ both functions agree. 
Similar behavior was observed for a film geometry \cite{SCHU1}. 

In order to relate our results to experimental results
let us now compute the finite-size scaling function $f_{1}(x)$ obtained
for bars with staggered boundary conditions in physical units. 
This leads to the same conversion formula used in Ref. \cite{SCHU1}:
$\left.f_{1}(x) \right|_{phys} = \lambda \left.f_{1}(x)\right|_{lattice}$
where the factor $\lambda$ is given by $\lambda = (V_{m}k_{B})/a^{3}  
({\mbox{\AA}}/a)^{\alpha/\nu}$ 
where $V_{m}$ is the molar volume of $^{4}He$ at saturated vapor pressure
at $T_{\lambda}$.  The unit of length $a$ (i.e., the lattice spacing $a$ 
in the XY model) was determined to be $a =2.95\mbox{\AA}$ 
\cite{SCHU1,SCHU2} and thus 
$\lambda  = 15.02 Joule/(^{\circ} K mole)$.

\begin{figure}[htp] 
 \centerline{\psfig {figure=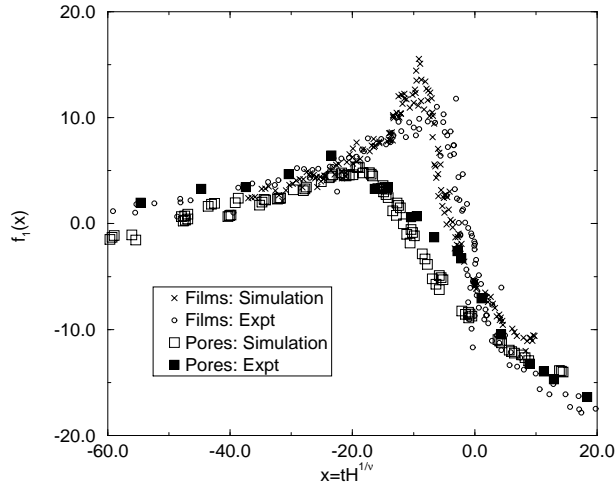,width=\mywidth}}
 \caption{\label{fig4} Our results for the specific heat scaling function
$f_{1}(x)$ for  pores (bars) with staggered boundary conditions  on the
substrate layers (open squares) are compared to the experimental 
results\protect\cite{COLEMAN} (solid squares). 
For comparison the same function is also
shown for the film geometry as obtained from our simulation (crosses,
taken from Ref. \protect\cite{SCHU1}) and from
experiments (open circles, taken from Ref. \protect\cite{LIPA}). 
$H$ is expressed in units of \mbox{\AA} and $f_{1}(x)$ in
$Joule/(^{\circ} K mole)$.}
\end{figure}

Coleman and Lipa \cite{COLEMAN} have recently measured the specific heat
of superfluid helium confined in pore geometry. They
have compared their results to the 
early data of Refs.\cite{EARLYC,CHEN}. There is some range of
disagreement near the critical temperature. The experimental results 
of Refs.\cite{EARLYC,CHEN}
have been obtained on much smaller diameter pores  than those of 
Ref.\cite{COLEMAN}. In our recent work for the effect of the boundaries on
superfluid films\cite{SCHU3} we found that the boundary can create
corrections to scaling which for films of thickness  as large as those of
the pore diameters of Refs.\cite{EARLYC,CHEN} can not be neglected. 
Since we want to stay
away from such difficulties in this paper, we decided to use the results
of the most recent work of Ref.\cite{COLEMAN} to compare with the
results of our theoretical calculations.
In Fig.~\ref{fig4} we compare the experimental results reported in
Ref.\cite{COLEMAN} with our Monte Carlo results. Since
the authors of Ref.\cite{COLEMAN} present only their specific heat data we
deduced the scaling function $f_{1}(x)$ according to Eq.(\ref{cimph}) using
the bulk data of liquid $^4$He reported in Ref.\cite{LIPA3}. The 
agreement between the
theoretical calculation and experiment is satisfactory. Note that the agreement
between the function $f_1$ obtained from Monte Carlo data in the presence
of {\it periodic} boundary conditions in the $H$-direction and the 
experimentally
determined function $f_1$ would be far worse. For example, the peak of the 
function $f_1(x)$ for {\it periodic} boundary conditions
in the $H$-direction in physical units is about $15 Joule/(^{\circ}Kmole)$ at 
$x \approx -4$, whereas for the experimental 
function $f_1(x)$ is approximately $6.4Joule/(^{\circ}Kmole))$ at $x
\approx -23.1$ (cf. Fig.~\ref{fig4}). The small difference between
experimental and theoretical scaling functions obtained with staggered
boundary conditions could be due to a number
of reasons, including poor representation of the geometry or the
boundary conditions in the real experimental system by the theoretical
modeling. 

Fig.~\ref{fig4} demonstrates again the effect of the type of the
confining geometry on the shape of the universal scaling function $f_{1}(x)$,
now in the presence of staggered boundary conditions which are a good
approximation to the physical boundary conditions imposed by the 
confining walls.

In conclusion, we have numerically computed the finite--size scaling
function $f_{1}(x)$ for a bar geometry and we have compared our results
to the scaling functions obtained earlier for cubes and films.
For pores (represented by bars) we find good agreement between our Monte Carlo
results in the presence of staggered boundary conditions  and the
most recent experimental results of Coleman and Lipa\cite{COLEMAN}. 
More precise experimental data for the 
specific heat of liquid $^4$He confined in pores near the $\lambda$-critical
temperature $T_{\lambda}$ are expected from future experiments\cite{LIPA2}
which can be compared to our calculation. These experiments could 
employ different substrates to check the influence of the boundary conditions
on the scaling function $f_1(x)$. Thus, this work together with our
earlier work on other geometry\cite{SCHU1,SCHU3} shows that only
scaling functions which belong to the same geometry and the same 
boundary conditions should be compared.

N.S. would like to thank
the H\"ochst\-leistungs\-rechen\-zen\-trum J\"ulich 
for the opportunity of using the Cray Y-MP.
This work was in part supported by Sonderforschungsbereich 341 der 
Deutschen Forschungsgemeinschaft and the National Aeronautics and Space
Administration under grant no. NAG3-1841.

\end{document}